# Fault-Resilient PCIe Bus with Real-time Error Detection and Correction

Mostafa Darvishi, Ph.D., *Member, IEEE*,
Independent High-Performance Computing Researcher

ABSTRACT

This paper presents a novel IP design for real-time fault/error detection and recovery on a peripheral component interconnect express (PCIe) which interfaces a host system (here a PC) to a slave design including processing system and memory transaction implemented on a Zynq Ultrascale Xilinx Kintex FPGA board (KCU105). The proposed IP design is capable of detection and correction of different types of PCIe errors on-the-fly.

*Index Terms*—PCI-Express, Xilinx Kintex Ultrascale FPGA, Error detection and correction.

## I. INTRODUCTION

FIELD Programmable Gate Arrays (FPGAs) are an attractive solution to implement systems for high-performance computing applications. Amongst different protocols to interface a host system (e.g., a computer) to a slave high-speed design (e.g., an FPGA design), peripheral component interconnect express (PCIe) is dedicated to interface high-speed components which was utilized for various applications in the past decade [1-5]. PCIe slots are available in different physical configurations, i.e., $x1$, $x4$, $x8$, $x16$, $x32$. The number after the $x$ determines how many lanes (how data travels to and from the PCIe card) that PCIe slot has. A PCIe $x1$ slot has one lane and can move data at one bit per cycle. A PCIe $x2$ slot has two lanes and can move data at two bits per cycle [6, 7]. Similar to any high-speed digital data transmission protocol, PCIe is not excepted from error bit either in transmitter (TX) or the receiver (RX) ends [8].

Several techniques have been proposed to test the vulnerability of PCIe in the literature [9-18]. A versatile hardware MitM architecture capable of interfacing with PCIe bus communications was presented in [9] which is amenable to a small range of applications. Most of the error detection techniques for PCIe reply on the off-line and delayed correction algorithms based on the software algorithms. The number of clock cycles to process the error detection algorithms will significantly affect the optimum slack for error correction cycle. The drawback of the technique presented in [18-18] is the application dependency and exhaustive work to find the error and also inability of error detection and correction in almost real-time mode. Moreover, increasing the design overhead will significantly affect the performance and usability of the proposed techniques.

In a recent proposed method called the "Jintide", which is especially suitable to constitute distributed large-scale clusters in CPU, it can amortize operation overheads. This scheme is effective in detecting pervasive hardware security issues, including vulnerabilities, backdoors, and hardware Trojans. However, the implementation overhead of the algorithm itself does not make it adoptable to multi-core CPU systems such as Zynq Ultrascale SoCs including real-time and application ARM Cortex cores.

The main contribution of this paper is evaluating the sensitivity of PCIe bus to different types of errors by an in-situ error detection and correction mechanism. The system also benefits a real-time fault injection core for testability and evaluation purposes. The injected error is stored in a DDR memory module which is continuously accessible by the Zynq processing system, software processor (MicroBlaze), and the host computer. The data discrepancy between the Zynq processor read-out mechanism and the host computer (via PCIe) is handled for correction by MicroBlaze and the original non-faulty data will be recovered at the expense of only one clock cycle. Instead of employing several error monitoring algorithms configured with algorithms, using a novel integrated IP for error detection and correction on-the-fly will allow the whole system to continue operating with no data transmission delay.

Several experiments were performed to identify different types of errors which are common in PCIe bus as well as performing a in-situ correction mechanism. Due to the page limitation of this paper, the results will be presented only for a few types of errors and the extended list of results will be drawn later at the time of presentation.

This paper is structured as follows. Different types of common errors in PCIe bus are classified and explained in Section II. This section is followed by motivation and methodology of employing an in-situ error detection and correction mechanism presented in Section III. Experiments and obtained results performed on a Zynq Ultrascale SoC FPGA is presented in Section IV and finally, we conclude in Section V.

## II. PCIe Errors Classification

In high-speed systems design, PCI Express has become the backbone. PCIe is a third-generation high performance I/O bus which is used to interconnect peripheral devices in applications such as computing, and communication platforms specifically tailored to high-speed applications. It is used to provide the connections between motherboard peripherals like graphics card, Ethernet card to the CPU and main memory.

Investigations on PCIe error handling on SoC devices has become crucial part because of application dependency. PCIe provides rich set of mechanisms for error recognition and handling where error handling may involve only hardware, device-specific software, or even the system software [19]. This paper describes the errors associated with the PCIe interface and error occurred while delivery of transactions between transmitter (host computer) and receiver (design implemented on the Zynq Ultrascale FPGA). Details of errors associated with each layer of PCIe, advanced error reporting (AER), advisory errors and recommendations for multiple error handling are described as follows.

### A. PCIe Errors Associated to Each Layer

PCIe is a packet-based serial bus which provides a secure channel for interconnecting high-speed devices together [19] while ensuring a high-speed, high-performance, point-to-point [18], dual simplex, and differential signaling link [16] between a source point (in our design, the host computer) and a decontamination point (in our design the targeted Zynq Ultrascale implementation). As shown in Fig. 1, PCIe has three layered architecture for communication between the source and destination device points, namely, "Transaction layer". "Data Link layer", and the "Physical layer". The errors associated to each architectural layer are described as follows [15-19].

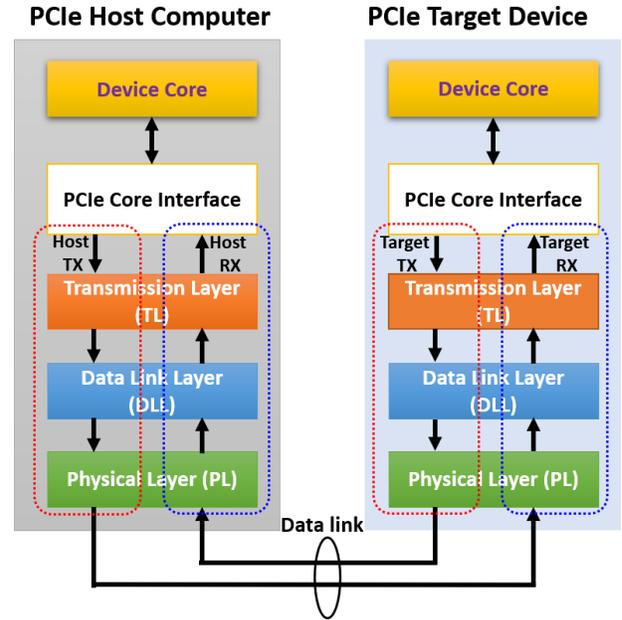

Fig. 1. PCIe bus layers architecture established between a host and a target device. In this paper, the host device is a computer, and the target device is a Zynq Ultrascale FPGA evaluation board (Xilinx KCU105).

### 1. Transaction Layer Errors

Transaction layer (TL) is the first and upper layer where the packet is formed. The transaction layer only checks the completion of data transfer for end-to-end device interconnections for the occurrence of the following errors:

- ECRC failure,
- Corrupted TLP, i.e., error occurred in packet format,
- Time-outs failure during separate packet transaction
- Flow control protocol error

- Unsupported requests for data transaction
- Data corruption, i.e., affected packets,
- Host system abort,
- Unexpected transaction completion: i.e., the slave handshakes for a data receive completion while the transmitter is still transferring data remained from the same packet,
- Receiver slave device overflow: i.e., the receiver stack will fill before ending the transaction sequence and the slave does not handshake for the transaction completion.

### 2. Data Link Layer Errors

Data Link layer (DLL) is the middle layer responsible for packet error and response handling. DLL will check the occurrence of the following errors in requester, switch link and the completer:

- LCRC failure for TL packets
- Sequence number check for TL packets
- LCRC failure for DLL packets
- Time-outs
- DLL protocol error

### 3. Physical Layer Errors

Physical layer (PL) is the third layer which is responsible for link training and transaction handling at interface level. PL will check the occurrence of the following errors in requester, switch link and the completer:

- Receiver errors, i.e., where the receiver reports or includes any receival of incomplete or corrupted packet due to an error,
- Link errors, i.e., the receiver includes the corrupted received packet due to a broken or affected link between the transmitter and receiver or even between the layer links.

### B. Severity of PCIe Errors

Depending on the severity of the PCIe errors and how they affect the data transaction between a source point and a destination point, the errors can be classified as follows [11-17]:

### 1. Correctible Errors

Correctible errors are addressed as those errors which impact the performance of the data transaction between a source point and a destination point such as bandwidth reduction or transmission latency. Correctible errors do not impact the data packet and data is not lost. The PCIe hardware will remain reliable and functional with the occurrence of correctible errors. Correctible errors will be handled and fixed by the hardware while no software intervention is required. Bad DLL packet is one of the correctible errors which is handled by the DL layer itself.

### 1. Non-correctible Errors

Non-correctible errors are considered as "*fatal*" or "*non-fatal*" error. In the event of fatal errors occurrence, the system software intervention is required to handle the error and rewrite the packet content fully or partially to repair the data. However, the non-fatal errors occurrence could be fixed by the intervention of a device-specific software written by the user. It is noted that in term of the overhead, system software intervention will be heavier than user-specified software intervention. The number of clock cycles required to repair a fatal error might be significantly higher than those needed for a non-fatal error repair on the same host and target device. Depending on the host computer and target device performances, the number of clock cycles for a fatal error repair could be an order of magnitude to the ones needed to fix a non-fatal error.

*Non-correctible fatal* errors will impact the integrity of the PCIe hardware established between the host (in our design the host computer) and the target device (in our design the Zynq Ultrascale FPGA). The PCIe data transmission link will be unstable, and data is lost in such event. That is the reason why the software system needs to intervene to handle this type of errors. The system software will handle and repair the non-correctible fatal errors by restarting both the target device and the PCIe link. Malformed TLP Error [16], Link Training Error[11], DLL Protocol Error [19], Receiver Overflow [17], and Flow Control

Protocol Error [14] are the examples of non-correctible fatal errors. In this paper, we will present how this type of error is handled by our proposed IP design implemented on the target device, i.e., the Zynq Ultrascale FPGA without necessity to restart the PCIe link and the target component. In our proposed IP design, the PCIe link will be recovered as well as the non-correctible fatal error on-the-fly thanks to the Partial Reconfiguration (PR) feature added to the IP. Details of this procedure will be resented in the following sections.

*Non-correctible non-fatal* errors will not impact the integrity of the PCIe hardware established between the host (in our design the host computer) and the target device (in our design the Zynq Ultrascale FPGA). However, the data is lost due to the occurrence of these errors. Non-correctible non-fatal errors will corrupt the content of data packet which cannot be repaired by the PCIe link at the hardware level. It is noted, though the occurrence of non-correctible non-fatal errors, the PCIe will remain reliable and fully functional and would continue to operate properly. In this case, the consecutive data transactions will be successful and safe but only a partial data in the precedent packet will be affected. Recovery of non-correctible non-fatal errors will be handled by the intervention of a user-specified software algorithm which will initiate a new transaction by the requester. Corrupted received TL packet [12], Unsupported Request (UR) [14, 19], Completion Timeout (CTO) [15], Completer Abort (CA) [13, 19], and Unexpected Completion (UC) [17-19] are the examples of non-correctible non-fatal errors. In this paper, we will discuss how this type of error is handled by our proposed IP design implemented on the target device, i.e., the Zynq Ultrascale FPGA which includes a specific user-defined software algorithm. As a conclusion of the PCIe errors classifications in this section, Table I summarizes different types of errors associated to he PCIe bus, their severity, a well as the examples of each error type and the corresponding PCIe layer which the error occurs [11, 19].

## III. PROPOSED IP DESIGN FOR PCIe ERROR DETECTION AND CORRECTION

This section presents a novel IP design for detection and correction of errors occurred in PCIe bus which links a source point to a destination point. The sources point in this paper will be a host computer running with a Linux Ubuntu 20.04 OS with x8 Gen 3 PCIe slot. The destination point, also called the target device, is a Xilinx Zynq Ultrascale SoC FPGA evaluation board (KCU105) which benefits from a x8 Gen 3 PCIe Core and associated interfacing pinouts. Fig. 2 shows the host computer (top) and targeted FPGA board (bottom) used in this paper with their respective PCIe slot and pinouts. The targeted FPGA board includes a bitstream of a design implemented by Xilinx Vivado 2019.2.

TABLE I. ERROR PES ASSOCIATED TO PCIE BUS WITH EXAMPLES AND POINT OF OCCURRENCE

| Type of error | Severity | Example | Corresponding PCIe layer which the error occurs |
|---|---|---|---|
| Correctible | Low | RX error (host& target) | PL |
| Correctible | Low | Bad TL packet | DL |
| Correctible | Low | Bad DLL packet | DL |
| Correctible | Low | Time-out | DL |
| Non-correctible non-Fatal | Medium | Corrupted RX TL packet | TL |
| Non-correctible non-Fatal | Medium | ECRC failure | TL |
| Non-correctible non-Fatal | Medium | Unsupported request | TL |
| Non-correctible non-Fatal | Medium | Completion time-out | TL |
| Non-correctible non-Fatal | Medium | Completion Abort | TL |
| Non-correctible non-Fatal | Medium | Unexpected Completion | TL |
| Non-correctible Fatal | High | Training error | PL |
| Non-correctible Fatal | High | DLL protocol error | DL |
| Non-correctible Fatal | High | RX overflow | TL |
| Non-correctible Fatal | High | Flow control protocol error | TL |
| Non-correctible Fatal | High | Corrupted TL packet | TL |

Fig. 3 shows the main block design of the proposed method for on-the-fly error detection and correction for h PCIe bus linking the host computer to the targeted KCU105 FPGA board. The main

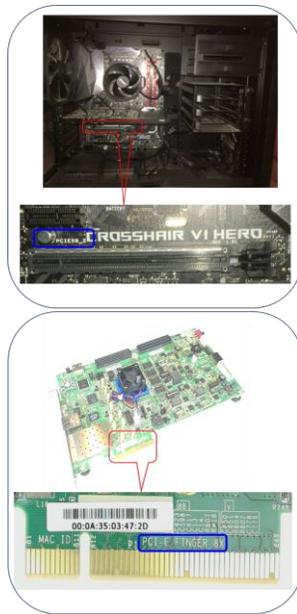

Fig. 2. Host computer (top) and targeted FPGA board (bottom), i.e., Xilinx Kintex Ultrascale KCU105, used in this paper with their respective PCIe slot and pinouts.

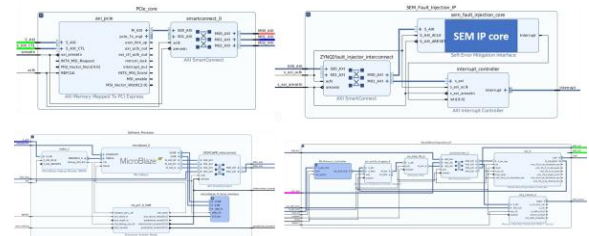

Fig. 4. Optimized demonstration of IP blocks defined for the design.

block design is comprised of the following IP cores:

- **PCIe core**; which is responsible to establish the PCIe link between the source point (host computer) and the destination (ZCU105 FPGA board);

- **Software Processor**; comprised of a MicroBlaze soft processor and corresponding memory modules for data transactions from/to the processor and its inbound and outbound modules (Fig. 3);

- **SEM Fault Injection IP**; comprised of a user-defined fault injection tool including the SEM IP core from Xilinx Inc. as well as the interrupt controller module to trigger the interrupt port of the Software Processor to start detecting discrepancies in data packets and then command the partial reconfiguration IP to start system reconfiguration on-the-fly. It is noted that this IP module is optional and is inserted in the design just for error injection and testing of the error detection mechanism. Indeed, the PCIe errors could also be generated through the channel itself.

- **ZYNQ Processor**; is the main processing system hardcoded inside the Kintex Ultrascale FPGA and is responsible for processing of the whole system and control of its slave modules. The Zynq Processor is implemented on the PS side of the FPGA fabric and includes multi-core ARM Cortex application and real-time microcontroller cores.

- **Partial Reconfiguration IP**; comprised of a Xilinx partial reconfiguration controller; DDR memory module, and a user-defined partial reconfiguration memory controller. This module as its name indicates, will perform the partial reconfiguration on the corrupted data packets on-the-fly while both host and destination devices continue to operate without system interruption. For the sake of space and limited number of pages, Fig. 4 shows only an optimized demonstration of IPs used in this design.

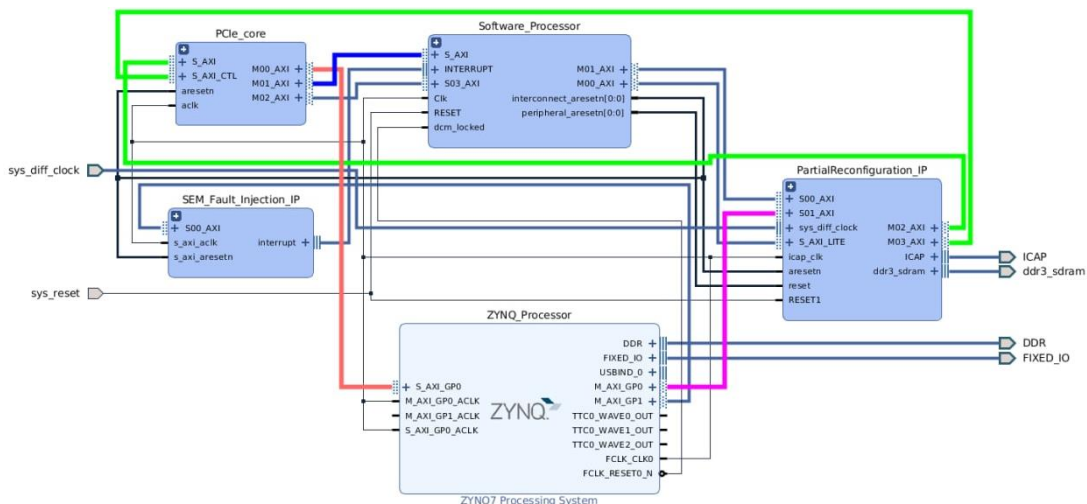

Fig. 3. Block design of the proposed technique implemented on the targeted CU105 FPGA board.

Fig. 5. Optimized demonstration of IP blocks defined for the design.

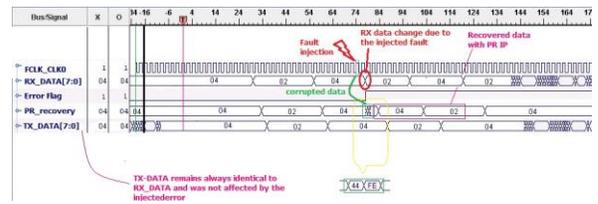

Fig. 6. Patrial result of error detection and correction for the PCIe bus. This is a TL-related error.

## IV. EXPERIMENTS AND RESULTS

The design in Fig. 3 was implemented on the Xilinx KCU105 development board connected to the PCIe x8 lane of the host computer (Fig. 2). After restarting the host computer, the KCU105 board will be recognized by the host computer via PCIe bus as shown in Fig. 5. Integrated Logic Analyzer (ILA) core was used to capture the corresponding PCIe bus signals for error detection. The SEM IP core which is an optional IP as described in Section III is only used for fault injection and quick testing of the fault detection scheme. The established PCIe bus between the host computer and the targeted device does not need this IP. The PCIe-related errors are usually generated during data transaction in different layers of the PCIe bus as described in Table I. The extended version of this paper with extensively discuss the root cause of errors how the proposed scheme will detect and correct them on-the-fly.

As shown in Fig. 3, the input system clock (`sys_diff_clock`) is an onboard 100 MHz clock fed to the partial reconfiguration IP which is also used by ILA to capture data signals at desired checkpoints. The rest of the system is clock with a 50 MHz clock (`FCLK_CLK0`) driven by the ZYNQ processing system. Clocking the entire system only with the single 100 MHz clock has two drawbacks; first, the distribution of the onboard 100 MHz clock signal created clock jitter for some modules, and, second, clocking the rest of the system with higher frequency does not add any value to the overall performance of the system. The `FCLK_CLK0` clock signal is a pure and jitter-free clock. The system reset is also provided through an onboard reset pin (`reset`).

*Design operating mechanism*: upon running the design, the fault injection tool (SEM) will start injecting faults (optional module). At the same time ILA is continuously capturing data transmitted to the PCIe bus and snapshots of data are being stored into the DDR3 memory located inside the partial reconfiguration module (see Fig 4). The stored data and the status of RX and TX signals from/to the PCIe are monitored by ZYNQ processor. Upon detection of a difference between RX and TX data packs, the error detection flag is raised. This flag sends an interrupt signal to the Software processor module, i.e., the MicroBlaze (pin interrupt from SEM_Fault_Injection _IP to Software_Processor IP in Fig. 3). The Software Processor module will then initialize the Partial Reconfiguration Controller core (Fig. 4) to start correcting the affected partial data packet. The recovered data will be captured again in the memory and is sent to the PCIe bus.

Fig. 6 shows a partial result for error detection in the PCIe bus and how the partial reconfiguration module (`PR_recovery` signal) corrects the erroneous data due to the injected fault. It is noted that the transmitted data (`TX_DATA`) always remains identical to the received data (`RX_DATA`) and is never informed about the occurred error because the error was masked and fixed on-the-fly. This type of error was a transaction layer (TL) error. For the sake of space in this paper, we could not cover detection of all other PCIe-related errors (Table I). Also, we avoided presenting the error flag circuitry presentation. Indeed, each PCIe error flag has a user-defined combinational circuity. More details will be presented at the time of the conference.

## V. CONCLUSION

This paper presented a novel IP design for resilient PCIe bus linking a host computer to a targeted Kintex Ultrascale FPGA device. The proposed IP design is capable of detection and correction of different types of PCIe errors on-the-fly. Future works include addressing all types of errors extensively for the PCIe bus as well as scheduling an experiment for fault injection at TRIUMF laboratory. These experiments are expected to be presented for the potential journal paper.